\documentclass[aps,pre,twocolumn,showpacs,twoside,floatfix,superscriptaddress]{revtex4}

\usepackage{graphicx}
\usepackage{graphics}
\usepackage{amsmath}
\usepackage{amssymb}
\usepackage{mathrsfs}
\usepackage{epsf}

\begin{document}

\title{Generalization of escape rate from a metastable state
driven by external cross-correlated noise processes}

\author{Jyotipratim Ray Chaudhuri}
\email{jprc_8@yahoo.com}
\affiliation{Department of Physics, Katwa College, Katwa,
Burdwan 713130, India}

\author{Sudip Chattopadhyay}
\email{sudip_chattopadhyay@rediffmail.com}
\affiliation{Department of Chemistry, Bengal Engineering and Science University, Shibpur,
Howrah 711103, India}

\author{Suman Kumar Banik}
\email{skbanik@phys.vt.edu} \affiliation{Department of Physics,
Virginia Polytechnic Institute and State University, Blacksburg, VA
24061-0435, USA}

\date{\today}

\begin{abstract}
We propose generalization of escape rate from a metastable state for
externally driven correlated noise processes in one dimension. In
addition to the internal non-Markovian thermal fluctuations, the
external correlated noise processes we consider are Gaussian,
stationary in nature and are of Ornstein-Uhlenbeck type. Based on a
Fokker-Planck description of the effective noise processes with
finite memory we derive the generalized escape rate from a
metastable state in the moderate to large damping limit and
investigate the effect of degree of correlation on the resulting
rate. Comparison of the theoretical expression with numerical
simulation gives a satisfactory agreement and shows that by
increasing the degree of external noise correlation one can enhance
the escape rate through the dressed effective noise strength.
\end{abstract}


\maketitle

\section{Introduction}

The theory of fluctuation induced barrier crossing dynamics was
first discussed in the seminal paper by Kramers \cite{hak} where he
considered a thermalized Brownian particle trapped in a one
dimensional well separated by a barrier of finite height from a
deeper well. The particle was supposed to interact with the
environmental degrees of freedom that act as a thermal reservoir.
The thermal environment exerts a damping force on the particle but
simultaneously thermally activates it so that the particle
effectively gains enough energy to cross the barrier of finite
height. Since its inception, Kramers' model has been extensively
revisited both theoretically \cite{htb,vim,ep} and experimentally
\cite{as,ewgd,lim} constituting a vast body of literature
\cite{uw,an}.

One of the essential conditions of the traditional approach to study
activated rate processes is the maintenance of a balance between the
two opposing forces, the thermal fluctuations and the dissipation,
which the Brownian particle experiences while in contact with the
thermal bath. Since these two counter balancing forces have a common
origin, the heat bath, it can be shown easily that they are
connected by fluctuation-dissipation relation (FDR) \cite{rk}. A
typical signature of FDR is that in the long time limit the Brownian
particle attains an equilibrium Boltzmann distribution for an
initial canonical distribution of the bath's degrees of freedom
\cite{rz}. In nonequilibrium statistical mechanical terminology,
such systems are referred to as closed system \cite{kl-bjw}. It may
happen sometime that an additional source of energy in the form of
fluctuations can be pumped from outside for which there is no
counter balancing force like dissipation \cite{wh-rl,jrc1,skb,jrc2}.
In absence of FDR, such external input of energy makes the system
open and in contrast to the closed system the equilibrium Boltzmann
distribution gets replaced by a steady state distribution (SSD) in
the long time limit. It may therefore be anticipated that the
absence of FDR tends to make the SSD function dependent on the
strength and correlation time of external noise as well as on the
dissipation of the system \cite{jrc1,skb,jrc2}. It is pertinent to
point out that though thermodynamically closed systems with
homogeneous boundary conditions possess, in general, time-dependent
solution, the driven open systems may settle down to complicated
multiple steady states when one takes into account nonlinearity of
the systems.

The origin of the noise in the thermodynamically open system driven
by two or more random forces may be different.  The barrier crossing
dynamics with multiplicative and additive noises arose strong
interest in the early eighties. The noise forces which appeared in
the dynamical system were usually treated as random variables
uncorrelated with each other. However, there are situations where
noises in some open systems may have common origin. If this happens
then the statistical properties of the noises should not be very
much different and can be correlated to each other. The
cross-correlated noises were first considered by Fedchenia
\cite{ref8} in the context of hydrodynamics of vortex flow. The
interference of additive and multiplicative white noises in the
kinetics of the bistable systems was analyzed by Fulinski and
Telejko \cite{ref9}. Maduriera {\em et al.} \cite{ref10} have
pointed out the probability of cross-correlated noise in ballast
resistor model showing bistable behavior of the system. Mei {\em et
al.} \cite{ref11} have studied the effects of correlations between
additive and multiplicative noise on the relaxation of the bi-stable
system driven by cross-correlated noises. It is now well accepted
that the effect of correlation between additive and multiplicative
noise is indispensable in explaining phenomena like phase
transition, transport of motor proteins, etc \cite{ref12}. As the
presence of the cross-correlated noises changes the dynamics of the
system \cite{ref13}, it is expected that there may exist some
additional effect of cross-correlation on the barrier-crossing
dynamics. The study of this additional effect acted precisely as the
catalyst that triggered the present study. From our formal
development and corresponding numerical application, it will be
revealed that the strength of correlation of noise process has
pronounced effect on the behavior of the escape rate of barrier
crossing process.

We are now in set to discuss the physical motivation of our model.
One can think about a system, simultaneously coupled to two
different heat baths. The two heat baths are in turn driven
externally by a random force $\epsilon(t)$ (say). The response of
$\epsilon(t)$ to the two baths are different. Now, if the
system-bath coupling be linear, apart from thermal noises due to the
presence of heat baths, the system will encounter {\it two additive
dressed noises} (dressed due to the bath-external noise coupling).
As both the baths are modulated by same noise with different
response, the dressed noises will be mutually correlated. Another
simple model can be invoked by considering the following
system-reservoir Hamiltonian: $H=H_{\rm S} + H_{\rm SB} + H_{\rm
int}$, where $H_{\rm S} = p^2/2 + V(x)$ is the system Hamiltonian
and $H_{\rm SB} = \sum_j [ p_j^2 + \omega_j^2 (q_j-c_j x)^2 ]/2 $
being the bath Hamiltonian (with interaction term between the system
and the heat bath), where the heat bath is assumed to be consisting
of harmonic oscillators with characteristic frequency set $\{
\omega_j \}$ as well as $\{ q_j,p_j\}$ are bath variables, and
$H_{\rm int} = \sum_j \left[ \kappa_j q_j \epsilon(t) + \lambda_j
p_j \epsilon(t)\right ]$. While writing the Hamiltonian we have
considered unit mass of the system and the bath oscillators. In this
model, initially the bath is in thermal equilibrium at the
temperature $T$ in the presence of the system, and then is
externally modulated by a noise $\epsilon(t)$. The coupling between
the bath and the noise is such that it linearly excites both
positions $q_j$ and momenta $p_j$ of the bath with the different
coupling constants $\kappa_j$ and $\lambda_j$ respectively. If we
now construct the equation of motion for the system variable, it can
be easily shown that the equation will contain, apart from the
thermal noise, two additive mutually correlated noises. As a more
physically motivated example, one can consider a typical
photochemical reaction where both reactant and solvent are exposed
to a weakly fluctuating light source. In such a case, the reactant
will be driven by two noises, first the external noise due to the
fluctuating light source and second, due to the presence of the
solvent which is also driven by the light source \cite{jrc2}. These
two noises will also be correlated as both the reactant and the
solvent are exposed to the same fluctuating light source. The above
mentioned physical examples lend firm support to our study presented
in this article.

The organization of the present work is as follows. In the next
section (Sec.~{\ref{model}}) we briefly describe the
phenomenological model for a system driven by cross-correlated
external fluctuations and characterize the statistical properties of
external driving. In Sec.~{\ref{escaperate}} we derive the
generalized escape rate from a metastable state and show different
limiting cases in moderate to large dissipation regime.
Computational details and results are provided in
Sec.{\ref{application}}. The paper is concluded in
Sec.~{\ref{conclusion}}.

%%%%%%%%%%%%%%%%%%%%%%%%%%%%%% SECTION II %%%%%%%%%%%%%%%%%%%%%%%%%%%%%
Before embarking on a discussion of the development and application
of our theory for escape rate from a metastable state driven by
external cross-correlated noise processes, in the next section, we
will discuss the essential ideas of our model. This will motivate us
towards the types of physically appealing approximations needed in
generating the required rate equations.

\section{The Model}\label{model}

To start with we consider the motion of a Brownian particle of unit
mass moving in an external force field $V(x)$. In the course of its
dynamics the particle experiences a random force $f(t)$ as well as a
counter balancing frictional force $\gamma(t)$ both originating from
the immediate thermal environment, the heat bath, to which the
particle is in contact. Since the aforesaid forces have a common
origin they are connected by the FDR (see Eq.(2) below) \cite{rk}.
Apart from the internal thermal noise, we assume that the particle
is driven by two external nonthermal, stationary, Gaussian,
Ornstein-Uhlenbeck noise processes $\epsilon(t)$ and $\pi(t)$ both
of which are correlated to each other by the correlation parameter
$\lambda$. The dynamics of the particle can then be described by the
generalized Langevin equation
\begin{equation}\label{eq1}
\ddot{x}+ \int_0^t \gamma(t-t^{\prime}) \dot{x}(t^{\prime})
dt^{\prime} + V^{\prime}(x) = f(t) + \epsilon(t) + \pi(t).
\end{equation}

\noindent Here the friction kernel $\gamma(t)$ is connected to the
internal noise $f(t)$ by the FDR of the second kind \cite{rk}
\begin{equation}\label{eq2}
\langle f(t) f(t^{\prime}) \rangle = k_BT \gamma(t-t^{\prime})
\end{equation}

\noindent where $T$ is the thermal equilibrium temperature and $k_B$
is the Boltzmann constant. The form of the Langevin equation
(\ref{eq1}) we have considered here guarantees that the nature of
the dynamics is of non-Markovian type and it is due to the finite
correlation effect of the thermal environment on the Brownian
particle. The nature of the dissipation kernel $\gamma (t)$ is very
much dependent on the nature of the coupling of the particle to the
heat bath and on the distribution of the bath modes \cite{rz}. In
the limit of vanishing correlation effect and for instantaneous
dissipation, i.e. for Markovian dynamics, Eq.(2) reduces to $\langle
f(t) f(t^{\prime}) \rangle = 2 \gamma k_BT \delta (t-t')$ with
$\gamma$ being the dissipation constant.

At this point it is pertinent to mention that by the application of
an external random force, fluctuations are created in a
deterministic system. As example, one may cite a noise generator
inserted into an electric circuit or a growth of species under the
influence of random weather. For such cases, the external
fluctuating force is never completely $\delta$-correlated or white.
On the other hand, the internal or intrinsic noise is caused by the
thermal fluctuations created due to the coupling with the
environmental degrees of freedom and cannot be completely switched
off. In general, the physical processes concerning chemical
reactions, growth of population etc. are of later type. For an
illuminating discussion in this context we refer to the book by van
Kampen \cite{vankampen}. At this point we note that a system where
internal noise is always present, can be driven by external noise
also. From a microscopic point of view, the above equations
(\ref{eq1}) and (\ref{eq2}) can be derived from a Hamiltonian, where
the system is coupled with a heat bath consisting of a set of
harmonic oscillators with different characteristic frequencies
\cite{rz} and is simultaneously driven by two external,
cross-correlated noises, $\epsilon(t)$ and $\pi(t)$. The system-bath
interaction generates the internal noise $f(t)$, statistical
properties of which will depend on the frequency spectrum of the
heat bath and on the nature of the system reservoir coupling. For a
finite correlation time $\tau_c$, (which can be considered as
inverse of the cut-off frequency of the bath), the internal noise
will be colored, while for $\tau_c \rightarrow 0$, the noise will be
$\delta$-correlated or white. Furthermore, the nonlinear
system-reservoir coupling yields multiplicative noise while bilinear
coupling gives additive noise. For further discussion, we refer to
the book of Lindenberg and West \cite{ kl-bjw}. In our present
study, all the noises have been assumed to be colored and additive.

The external noise processes we have considered are independent of
the dissipation kernel, hence there exist no corresponding FDR for
them. In addition to that, we also assume that the external noises
do not influence the internal noise process and hence $\epsilon(t)$
and $\pi(t)$ are statistically independent of $f(t)$. The physical
situation we address here is that the system is in thermal
equilibrium at $t=0$ in presence of the thermal bath but in absence
of the external noise processes. At $t=0+$, the external
fluctuations are switched on and the system is driven by the two
external noises $\epsilon(t)$ and $\pi(t)$ which are correlated to
each other. The system dynamics is then governed by the generalized
Langevin equation (\ref{eq1}). We characterize the statistical
properties of the correlated external fluctuations by the following
set of equations
\begin{subequations}
\begin{eqnarray}
\langle\epsilon(t)\rangle &=& \langle\pi(t)\rangle = 0 \\
\langle\epsilon(t)\epsilon(t^{\prime})\rangle &=&
\frac{D_{\epsilon}}{\tau_{\epsilon}}
\exp \left (-\frac{|t-t^{\prime}|}{\tau_{\epsilon}} \right ) \\
\langle\pi(t)\pi(t^{\prime})\rangle &=& \frac{D_{\pi}}{\tau_{\pi}}
\exp \left (-\frac{|t-t^{\prime}|}{\tau_{\pi}} \right ) \text{ and }
\\
\langle\epsilon(t)\pi(t^{\prime})\rangle &=& \langle
\pi(t)\epsilon(t^{\prime})\rangle \nonumber \\
&=& \frac { \lambda ( D_{\epsilon} D_{\pi} )^{1/2}} {\tau} \exp
\left (-\frac{|t-t^{\prime}|}{\tau} \right ). \label{eq3}
\end{eqnarray}
\end{subequations}

\noindent In the above equations, $D_{\epsilon}$ and
$\tau_{\epsilon}$ are the strength and correlation time of the noise
$\epsilon(t)$, while $D_{\pi}$ and $\tau_{\pi}$ correspond to the
noise $\pi{(t)}$. $\lambda$ denotes the degree of correlation
between the noise processes $\epsilon(t)$ and $\pi(t)$. At this
point, we define $\xi(t)$ [$\xi(t) = (\epsilon(t)+ \pi(t))$ ] as the
effective external noise process whose statistical properties can be
defined using the properties of $\epsilon(t)$ and $\pi(t)$. Our
following analysis will be based on this effective noise $\xi(t)$
and we shall study the dependence of physical quantities in terms of
this effective noise $\xi(t)$ and the effective correlation time.
Since both the noise processes are Gaussian and simultaneously
stationary with zero mean, $\xi(t)$ will also be stationary Gaussian
with zero mean. Furthermore, the second moment of $\xi(t)$ is given
by
\begin{equation}\label{eq4}
\langle \xi(t) \xi(t^{\prime}) \rangle = \frac{D_R}{\tau_R} \exp
\left (-\frac{|t-t^{\prime}|}{\tau_R} \right )
\end{equation}

\noindent where the strength $D_R$ and the correlation time $\tau_R$
of the effective noise can be written as
\begin{eqnarray} \label{eq5}
D_R &=& \int_0^\infty dt \langle \xi(t) \xi(0) \rangle =
D_{\epsilon} + 2 \lambda \sqrt{D_{\epsilon} D_{\pi}} +D_{\pi} \\
\tau_R & = & \frac{1}{D_R} \int_0^\infty t \langle \xi(t) \xi(0)
\rangle dt \nonumber \\
& = & \frac{1}{D_R} \left [ D_{\epsilon} \tau_{\epsilon} +  2
\lambda \sqrt{D_{\epsilon} D_{\pi}} \tau +D_{\pi} \tau_{\pi} \right
]. \label{eq6}
\end{eqnarray}

\noindent It is apparent from equations (\ref{eq4}-\ref{eq6}) that
variation of the degree of correlation $\lambda$ will change the
strength and correlation time of the effective noise.

From the above equations (\ref{eq4}-\ref{eq6}), it is clear that
because of the cross correlation the Brownian particle realizes
the effect of two external noise processes by the effective noise
process $\xi(t)$ with the effective noise strength $D_R$ and the
noise correlation time $\tau_R$ containing the noise strength and
correlation time of $\epsilon(t)$ and $\pi(t)$ and the degree of
noise correlation parameter $\lambda$.

%%%%%%%%%%%%%%%%%%%%%%%%%%%%% SECTION III %%%%%%%%%%%%%%%%%%%%%%%%%%%%%

\section{Generalization of escape rate}\label{escaperate}

The modifications of the standard escape rate from a metastable
state in presence of correlated fluctuations are twofold. First, in
presence of the external correlated fluctuations, the dynamics
around the barrier top gets affected in such a way that the
stationary flux across the barrier top gets modified. Second, the
equilibrium Boltzmann distribution at the source well gets replaced
by a steady state distribution reflecting the signature of extra
energy input due to the open system. With these major changes in the
dynamics we then derive the generalized escape rate and show various
limits of the rate expression in the following analysis.

To analyze the dynamics across the barrier top we first linearize
the potential $V(x)$ around the barrier top at $x \approx 0$
\begin{equation}\label{eq7}
V(x) = E_b - \frac{1}{2} \omega_b^2 x^2 + \cdots \text{ , }
\omega_b^2 > 0
\end{equation}

\noindent where $\omega_b$ is the linearized frequency at the barrier top and
$E_b$ ($=V(0)$) is the barrier height. Thus, the linearized version of the
Langevin equation (\ref{eq1}) takes the form
\begin{equation}\label{eq8}
\ddot{x}+ \int_0^t \gamma(t-t^{\prime}) \dot{x}(t^{\prime})
dt^{\prime} - \omega_b^2 x  = F(t)
\end{equation}

\noindent where $F(t)$ is defined as
\begin{equation}\label{eq9}
F(t) = f(t) + \xi(t).
\end{equation}

\noindent The general solution of Eq.(\ref{eq7}) is given by
\begin{equation}\label{eq10}
x(t) = \langle x(t) \rangle + \int_0^t M_b(t-t^{\prime}) F(t^{\prime})
dt^{\prime}
\end{equation}

\noindent where
\begin{equation}\label{eq11}
\langle x(t) \rangle = v_0 M_b(t) + x_0 \chi_x^b (t)
\end{equation}

\noindent with $x_0= x(0)$ and $v_0=\dot{x}(0)$ being the initial
position and velocity of the particle and
\begin{equation}\label{eq12}
\chi_x^b (t) = 1 + \omega_b^2 \int_0^t M_b(t^{\prime}) dt^{\prime}.
\end{equation}

\noindent The kernel $M_b(t)$ is the Laplace inverse of
\begin{equation}\label{eq13} \tilde{M}_b(s) = \frac{1}{s^2+s
\tilde{\gamma}(s) - \omega_b^2 }
\end{equation}
with $\tilde{\gamma}(s) = \int_0^\infty \exp(-st) \gamma(t) dt$, being
the Laplace transform of $\gamma(t)$.
The time derivative of Eq.(\ref{eq10}) gives
\begin{equation}\label{eq14}
v(t) = \langle v(t) \rangle + \int_0^t m_b(t-t^{\prime})
F(t^{\prime}) dt^{\prime}
\end{equation}

\noindent where
\begin{equation}\label{eq15}
\langle v(t) \rangle = v_0 m_b(t) + \omega_b^2 x_0 M_b(t)
\end{equation}

\noindent and
\begin{equation}\label{eq16}
m_b(t) = \frac{dM_b(t)}{dt}.
\end{equation}

\noindent As both the internal and external noise processes are
stationary, one can explicitly use the stationary property of these
fluctuations to write the correlation function of $F(t)$ as $\langle
F(t) F(t^{\prime})\rangle = c(t-t^\prime)$ along with the symmetry
condition $c(t-t^{\prime}) = c(t^{\prime}-t)$. Using this form of
stationarity we calculate the variances in terms of $M_b(t)$ and
$m_b(t)$ as
\begin{eqnarray}
\sigma_{xx}^2(t) & = & \langle [x(t)- \langle x(t)\rangle]^2\rangle
\nonumber \\
                 & = & 2 \int_0^t M_b(t_1) dt_1
                         \int_0^{t_1} M_b(t_2) c(t_1-t_2) dt_2 \label{eq18} \\
\sigma_{vv}^2(t) & = & \langle [v(t)- \langle v(t)\rangle]^2\rangle
\nonumber \\
                 & = & 2 \int_0^t m_b(t_1) dt_1
                     \int_0^{t_1} m_b(t_2) c(t_1-t_2) dt_2 \label{eq19} \\
\sigma_{xv}^2(t) & = & \langle [x(t)- \langle x(t)\rangle][v(t)-
\langle v(t)\rangle]\rangle
\nonumber \\
                 & = & \int_0^t m_b(t_1) dt_1
                       \int_0^t M_b(t_2) c(t_1-t_2) dt_2 \label{eq20}
\end{eqnarray}

\noindent From Eqs.(\ref{eq18}) and (\ref{eq20}) we observe that
\begin{equation}\label{eq21}
\sigma_{xv}^2(t) = \dot{\sigma}_{xx}^2(t)/2.
\end{equation}

\noindent Using the method of characteristic function \cite{jm-jmp}
we then write the Fokker-Planck equation for probability
distribution function $P(x,v,t)$ near the barrier top as
\begin{eqnarray}\label{eq22}
\frac{\partial P}{\partial t} & = & - v \frac{\partial P}{\partial
x} - \bar{\omega}_b^2(t) x \frac{\partial P}{\partial v} +
\bar{\gamma}_b(t) \frac{\partial (vP)}{\partial v} \nonumber \\
&& + \phi_b(t) \frac{\partial^2 P}{\partial v^2} + \psi_b(t)
\frac{\partial^2 P}{\partial x \partial v}
\end{eqnarray}

\noindent where the subscript `$b$' signifies the dynamical
quantities defined at the barrier top and they are given by
\begin{eqnarray}\nonumber
\bar{\gamma}_b(t) &=& - \frac {d}{dt} \ln Y_b(t) \\
\nonumber
\bar{\omega}_b^2(t) &=& \frac { - M_b(t) \dot{m}_b(t) + M_b^2(t)} {Y_b(t)} \\
\nonumber
Y_b(t) &=& - \frac {m_b(t)}{\omega_b^2} \left \{ 1+
\omega_b^2 \int_0^t M_b(t^{\prime}) dt^{\prime} \right \} + M_b^2(t) \\
\nonumber \phi_b(t) &=& \bar{\omega}_b^2(t) \sigma_{xx}^2(t) +
\bar{\gamma}_{b}(t) \sigma_{vv}^2(t) +\frac{1}{2} {\dot
\sigma}_{vv}^2(t) \\
\psi_b(t) &=& \bar{\omega}_b^2(t) \sigma_{xx}^2(t) +
\bar{\gamma}_{b}(t) \sigma_{xv}^2(t) +{\dot \sigma}^2_{xv}(t) -
\sigma_{vv}(t)^2. \nonumber \\
\label{eq23}
\end{eqnarray}

\noindent Although bounded, these time-dependent parameters may not
always provide long time limits. In general, one has to work with
frequency $\bar{\omega}_b(t)$ and friction $\bar{\gamma}_{b}(t)$ for
analytically tractable models. In absence of external noise
$\phi_b(t) = k_b T \bar{\gamma}_{b}(t) $ and $\psi_b(t) = (k_B T
/\omega_b^2) [\bar{\omega}_b^2 -{\omega}_b^2]$ and in the Markovian
limit $\bar{\gamma}_{b}(t) = \gamma \delta(t)$,
$\bar{\omega}_b = {\omega}_b$ and consequently, $\psi_b(t)$
vanishes.

To calculate the stationary distribution near the bottom of the left
well, we now linearize the potential $V(x)$ around $x \approx x_0$.
The corresponding Fokker-Planck equation describing the dynamics at
the source well can again be constructed using the method of
characteristic function,
\begin{eqnarray}
\frac{\partial P}{\partial t} &=& - v \frac{\partial P}{\partial x}
- \bar{\omega}_0^2(t) x \frac{\partial P}{\partial v} + \bar
{\gamma}_0(t) \frac{\partial (vP)}{\partial v} +
\phi_0(t) \frac{\partial^2 P}{\partial v^2} \nonumber \\
&& + \psi_0(t)
\frac{\partial^2 P}{\partial x \partial v} \label{eq24}
\end{eqnarray}

\noindent with
\begin{eqnarray}
\nonumber
\bar{\gamma}_0(t) &=& - \frac {d}{dt} \ln Y_0(t) \\
\nonumber
\bar{\omega}_0^2(t) &=& - \frac {M_0(t) \dot{m}_0(t) +
M_0^2(t)} {Y_0(t)} \\
\nonumber
Y_0(t) &=& - \frac {m_0(t)}{\omega_0^2} \left \{ 1+
\omega_0^2 \int_0^t M_0(t^{\prime}) dt^{\prime} \right \} + M_0^2(t) \\
\nonumber
\phi_0(t) &=& \bar{\omega}_0^2(t) \sigma_{xx}^2(t) +
\bar{\gamma}_0(t) \sigma_{vv}^2(t) +\frac{1}{2} {\dot
\sigma}_{vv}^2(t) \\
\label{eq25} \psi_0(t) &=& \bar{\omega}_0^2(t) \sigma_{xx}^2(t) +
\bar{\gamma}_0(t) \sigma_{xv}^2(t) +{\dot \sigma}_{xv}^2(t) -
\sigma_{vv}^2(t).\nonumber \\
\end{eqnarray}

\noindent Here, the subscript `0' signifies the dynamical quantities
corresponding to the bottom of the left well. As mentioned earlier,
in absence of external noise and in the Markovian limit, $\psi_0(t)$
vanishes with $\bar{\gamma}_0(t) = \gamma \delta(t)$,
$\phi_0(t)=\gamma k_BT$ and $\bar{\omega}_0=\omega_0$. As a result,
we recover the Fokker-Planck equation for a harmonic oscillator with
frequency $\omega_0$. Thus, we identify Eq.(\ref{eq24}) as the
generalized version of the non-Markovian Fokker-Planck equation for
a harmonic oscillator which is driven by two externally correlated
noise processes. At this juncture it is pertinent to mention the
fact that the form of our Eq.(\ref{eq24}) is exactly identical with
that of the non-Markovian Fokker-Planck equation for harmonic
oscillator derived earlier by Adelman \cite{saa}. The steady state
[$\partial P(x \approx x_0,v)/\partial t=0$] solution of
Eq.(\ref{eq24}) is given by
\begin{equation}\label{eq26}
P_{st}^0(x,v) = \frac{1}{Z} \exp \left [- \frac{v^2}{2D_0} -
\frac{\bar{\omega}_0 x^2}{D_0+\psi_0} \right ]
\end{equation}

\noindent where $D_0 = \phi_0/\bar{\gamma}_0$. The quantities
$\psi_0$, $\phi_0$ and $\bar{\gamma}_0$ are long time (steady-state)
limit of the time dependent functions $\psi_0(t)$, $\phi_0(t)$ and
$\bar{\gamma}_0(t)$ respectively and $Z$ is the normalization
constant. The above solution (\ref{eq26}) can be verified by
directly putting it into the steady state version ($\partial
P/\partial t=0$) of Eq.(\ref{eq24}). The distribution function
(\ref{eq26}) is the steady state counterpart of equilibrium
Boltzmann distribution ($\exp \{ -[v^2+V(x)]/k_BT \}$) for
nonequilibrium open system. In the limit of pure thermal processes,
i.e. in absence of external fluctuating driving force, it is easy to
recover the equilibrium Boltzmann distribution from (\ref{eq26}).
The justification for using the distribution (\ref{eq26}) is the
following. In the traditional theory of activated rate processes
within the framework of pure thermal fluctuations the equilibrium
Boltzmann distribution is necessary to initially thermalize the
reactant state \cite{hak,htb}. On the other hand, for nonequilibrium
open system, a constant input of energy through external fluctuating
driving force forbids the system to attain the equilibrium state,
hence the system approaches (as in our case) towards the steady
state (\ref{eq26}), an analogue of equilibrium state, to initially
energize the reactant state by the effective temperature like
quantity (a complex function of $\bar{\gamma}_0$, $\phi_0$ and
$\psi_0$) embedded in the distribution function (\ref{eq26}).

To calculate the stationary current across the barrier top (for pure
thermal processes) within the framework of Kramers' original
reasoning \cite{hak}, one considers an equilibrium Boltzmann
distribution multiplied by a propagator $G(x,v)$ and uses it to
solve the Fokker-Planck equation (\ref{eq22}). From the reasoning of
the previous paragraph, it is clear that in our analysis, the
equilibrium distribution should be replaced by a SSD that contains
all the information about the potential around the barrier top and
the effective temperature like quantity for nonequilibrium open
system. In the limit of pure thermal processes, i.e. in absence of
any external fluctuations the SSD should reduce to the equilibrium
Boltzmann distribution. Thus, in the light of Kramers' ansatz we
consider a solution of Eq.(\ref{eq22}) at the stationary limit as
\begin{equation}\label{eq27}
P_{st}^b (x,v) = \exp \left [- \frac{v^2}{2D_b} - \frac{ \tilde{V}(x)
}{D_b+\psi_b} \right ] G(x,v)
\end{equation}

\noindent where $D_b = \phi_b/\bar{\gamma}_b$ and $\psi_b$ are the
long time limits of the corresponding time-dependent quantities
calculated at the barrier top region and $\tilde{V}(x)$ is the
linearized potential near the barrier top ($x \approx 0$) with a
renormalization in its frequency,
\begin{equation}\label{eq28}
\tilde{V} (x) \simeq  E_b - \frac{1}{2} \bar{\omega}_b^2 x^2
\end{equation}

\noindent where $\bar{\omega}_b$ is the long limit of
$\bar{\omega}_b(t)$ and $E_b$ is the barrier height. In writing
Eq.(\ref{eq28}) it has been assumed that the position of the maxima
of the potential $V(x)$ and the barrier height remains unchanged
while considering the memory effect in the dynamics. {\it The
non-Markovian effects are reflected only in the frequency}. The
ansatz of the form (\ref{eq27}) denoting the SSD is motivated by the
local analysis near the source well and the top of the barrier in
the Kramers' sense. Inserting Eq.(\ref{eq27}) in Eq.(\ref{eq22}) we
get in the steady state ($\partial P/\partial t =0$) an equation for
the function $G(x,v)$ as
\begin{eqnarray}
&& - \left ( 1+\frac{\psi_b}{D_b} \right ) v \frac{\partial
G}{\partial x} - \left [\frac{D_b}{D_b+\psi_b} \bar{\omega}_b^2 x +
\bar{\gamma}_b v  \right ]\frac{\partial G}{\partial v}\nonumber \\
&& + \phi_b
\frac{\partial^2 G}{\partial v^2} + \psi_b \frac{\partial^2
G}{\partial x \partial v} =0. \label{eq29}
\end{eqnarray}

\noindent We now introduce a variable $u$ as
\begin{equation}\label{eq30}
u = v + a x
\end{equation}

\noindent where $a$ is a constant to be determined. With the help of
the transformation (\ref{eq30}), Eq.(\ref{eq29}) reduces to
\begin{eqnarray}
&& (\phi_b+ a \psi_b) \frac{d^2 G}{d u^2} - \left
[\frac{D_b}{D_b+\psi_b} \bar{\omega}_b^2 x \right. \nonumber \\
&& \left. + \left \{ \bar{\gamma}_b + a \left ( 1 +
\frac{\psi_b}{D_b} \right ) \right \} v \right ]\frac{d G}{du} = 0.
\label{eq31}
\end{eqnarray}

\noindent At this point, we define
\begin{equation}\label{eq32}
\frac {D_b}{D_b+\psi_b} \bar{\omega}_b^2 x + \left \{ \bar{\gamma}_b
+ a \left ( 1 + \frac{\psi_b}{D_b} \right )  \right \} v = - \mu u
\end{equation}

\noindent where $\mu$ is another constant to be determined. From
Eqs. (\ref{eq30}) and (\ref{eq32}), we find that the constant `$a$'
has two values,
\begin{equation}\label{eq33}
a_{\pm}  = - \frac{B}{2A} \pm \sqrt{\frac{B^2}{4A^2} + \frac{C}{A} }
\end{equation}

\noindent with
\begin{equation}
A = \left (1+ \frac{\psi_b}{D_b} \right ), B= \bar{\gamma}_b \text{
and } C= \frac {D_b}{(D_b+\psi_b)} \bar{\omega}_b^2. \label{eq34}
\nonumber
\end{equation}

\noindent With the help of Eq.(\ref{eq32}), Eq.(\ref{eq31}) can then
be written as
\begin{equation}\label{eq35}
\frac{d^2}{du^2}G + \Lambda \frac{d}{du}G =0,
\end{equation}

\noindent where
\begin{equation}\label{eq36}
\Lambda = \frac {\mu}{\phi_b+ a \psi_b}.
\end{equation}

\noindent The general solution of Eq.(\ref{eq35}) is
\begin{equation}\label{eq37}
G(u) = G_2 \int_0^u \exp \left ( -\frac{1}{2} \Lambda z^2 \right )
dz + G_1
\end{equation}

\noindent where $G_1$ and $G_2$ are the constants of integration.
We look for a solution which vanishes for large $x$. This
condition is satisfied if the integration in Eq.(\ref{eq37})
remains finite. It is easy to understand that the integral in
Eq.(\ref{eq37}) converges for $|u| \rightarrow \infty$ if
and only if $\Lambda$ is positive. The positivity of $\Lambda$
depends on the sign of `$a$' and we observe that the negative
value of `$a$', i.e., `$a_{-}$' guarantees the positive value of
$\Lambda$.

To determine the value of $G_1$ and $G_2$, we now demand that
$G(x,v) \rightarrow 0$ for $x \rightarrow +\infty $ and for all $v$.
This condition yields $G_1 = G_2 \sqrt{\pi/2 \Lambda}$, so that
\begin{equation}\label{eq38}
G(u) = G_2 \left [ \left ( \frac{\pi}{2 \Lambda} \right )^{1/2}+
\int_0^u \exp \left (-\frac{1}{2} \Lambda z^2 \right ) \right ].
\end{equation}

\noindent Consequently, the stationary solution near the barrier top
becomes
\begin{eqnarray}
&& P_{st}^b(x \approx 0, v) \nonumber \\
&& = G_2 \exp \left (
-\frac{V(0)}{D_b+\psi_b} \right ) \nonumber \\
&& \times \left [ \left ( \frac{\pi}{2 \Lambda} \right )^{1/2} \exp
\left ( -\frac{v^2}{2D_b} \right ) + g(x \approx 0,v) \exp \left
(-\frac{v^2}{2D_b} \right ) \right ] \nonumber \\
\label{eq39}
\end{eqnarray}

\noindent with
\begin{equation}\label{eq399}
g(x,v) = \int_0^{v+a_{-}x} \exp \left (- \frac{1}{2} \Lambda z^2 \right ) dz.
\end{equation}

\noindent Since the steady state current $j$ across the barrier is defined as
\begin{equation}\label{eq40}
j = \int_{-\infty}^{+\infty} v P_{st}^b (x \approx 0, v) dv,
\end{equation}

\noindent we get using Eq.(\ref{eq39})
\begin{equation}\label{eq41}
j = G_2 D_b \left ( \frac{2\pi}{\Lambda + D_b^{-1}} \right )^{1/2}
\exp \left (-\frac{E_b}{D_b+\psi_b} \right ).
\end{equation}

\noindent To obtain the remaining constant $G_2$, we note that as
$x\rightarrow -\infty$, the term $G_2 [ \sqrt{\pi/2 \Lambda} +g]$ in
Eq.(\ref{eq38}) reduces to $G_2 \sqrt{ \pi/2 \Lambda}$. We then
obtain the reduced distribution function \cite{rdf} in $x$ as
\begin{equation}\label{eq42}
{\tilde P}_{st}^b(x\rightarrow -\infty) = 2 \pi G_2 \left (
\frac{D_b}{\Lambda} \right )^{1/2} \exp \left (
-\frac{\tilde{V}(x)}{D_0+\psi_0} \right ).
\end{equation}

\noindent Similarly, we derive the reduced distribution function in
the left well around $x \approx x_0$ as
\begin{equation}\label{eq43}
{\tilde P}_{st}^0(x) =\frac{1}{z} \sqrt{2 \pi D_o}
\exp \left (-\frac{E_0}{D_0+\psi_0} \right )
\end{equation}

\noindent where we have used Eq.(\ref{eq26}) and employed the
expansion of ${\tilde V}(x)$ as
\begin{equation}\label{eq44}
{\tilde{V}}(x) \simeq E_0+ \frac{1}{2} \bar{\omega}_0^2 {(x-x_0)}^2
\text{ , } x \approx x_0.
\end{equation}

\noindent At this point, we impose another condition that at $x
\approx x_0$, the reduced distribution function (\ref{eq42}) must go
over to the stationary reduced distribution function (\ref{eq43}) at
the bottom of the left well. This matching condition
\cite{jrc1,skb,jrc2} along with the normalization condition,
$\int_{-\infty}^{+\infty} P_{st}^0(x,v) dx dv=1$ gives the value of
the remaining constant $G_2$ as
\begin{equation}\label{eq45}
G_2 = \left ( \frac {\Lambda}{D_b} \right )^{1/2} \frac
{\bar{\omega}_0}{ [8\pi^3 (D_0+\psi_0)]^{1/2}  }.
\end{equation}

\noindent Hence, from Eq.(\ref{eq41}), we get the expression for the
normalized current or barrier-crossing rate
\begin{equation}\label{eq46}
k = \frac {\bar{\omega}_0 }{2\pi} \frac{D_b}{(D_0+\psi_0)^{1/2}}
\left ( \frac{\Lambda}{1+\Lambda D_b} \right )^{1/2} \exp \left (-
\frac{E}{D_b+\psi_b} \right )
\end{equation}

\noindent where $E$ $(=E_b-E_0)$ is the activation energy. In
passing we note that the temperature due to internal thermal noise,
the strength of the external noise, the correlation times and the
damping constant, are all buried in the expression for the
generalized escape rate for the open system through the parameters
$D_0, D_b, \psi_0, \psi_b$ and $\Lambda$. We also note that $k$ is
also an implicit function of the degree of correlation $\lambda$.

From the structure of Eq.(\ref{eq46}) it is difficult to understand
the role of various parameters (internal or external) on the rate
expression. We thus consider different limiting cases in the
following part to see the behavior of the rate expression.

First, we consider the case with no external driving and the
internal thermal noise being $\delta$-correlated, i.e.
\begin{equation}\nonumber
\epsilon(t) = \pi(t) =0 \text{ and }
\langle f(t) f(t^{\prime}) = 2 \gamma k_B T \delta (t-t^{\prime}).
\end{equation}

\noindent Making use of the abbreviations in Eqs.(\ref{eq23}) and
(\ref{eq25}), one can show that for pure $\delta$-correlated thermal
process $\psi_0 = \psi_b =0$, $D_b = D_0 = k_B T $, $\Lambda =
\mu/(\gamma k_B T)$, $\mu = - (a_{-} + \gamma ) $, and $a_{-} = -
(\gamma/2) - [(\gamma^2/4) + \omega_b^2 ]^{1/2}$. Thus, the general
expression (\ref{eq46}) reduces to the classical expression for
Kramers rate
\begin{equation}\label{eq48}
k_{\rm Kramers} = \frac{\omega_0}{2 \pi \omega_b}
\left [ \left ( \frac {\gamma^2}
{4} + \omega_b^2 \right )^{1/2}  - \frac{\gamma}{2} \right ] \exp
\left (-\frac{E}{k_B T} \right ).
\end{equation}

Second, we consider the case with no external noise but the internal
dynamics is non-Markovian with an exponentially decaying memory
kernel. In this limit $\epsilon(t)= \pi(t) =0 $, $\langle f(t)
f(t^{\prime}) \rangle = ({\cal D}/\tau_c) \exp(-|t-t'|/\tau_c)$
where $\cal{D}$ denotes the noise strength $({\cal D}= 2 \gamma k_B
T)$ and $\tau_c$ refers to the correlation time of the internal
noise processes. Then, from Eqs.(\ref{eq23}) and (\ref{eq24}) we
obtain
\begin{eqnarray*}\nonumber
&& D_0 =D_b = k_B T \\
&& \psi_0 = k_B T \left (1 - \frac{ \bar{\omega}_0^2 } {\omega_0^2}
\right ),
\psi_b = k_B T \left ( 1 - \frac{ \bar{\omega}_b^2 } {\omega_b^2} \right ),\\
&& \mu = - \left [ \bar{\gamma}_b + \left (
\frac{ \bar{\omega}_b^2 } {\omega_b^2} \right )a_{-} \right ], \\
&& a_{-} = \frac {\omega_b^2} { \bar{\omega}_b^2 } \left [ -
\frac{\bar{\gamma}_b}{2} - \left ( \frac{\bar{\gamma}_b^2}{4} +
\bar{\omega}_b \right )^{1/2} \right ]
\end{eqnarray*}

\noindent and hence the rate becomes
\begin{equation}\label{eq49}
k_{\rm memory} = \frac{\omega_0}{2 \pi \omega_b} \left [ \left ( \frac
{\bar{\gamma}_b^2} {4} + \bar{\omega}_b^2 \right )^{1/2} -
\frac{\bar{\gamma}_b}{2} \right ] \exp \left ( -\frac{E}{k_B T}
\right ).
\end{equation}

\noindent Eq.(\ref{eq49}) is the rate expression for pure thermal
process with exponentially decaying memory kernel and was derived
several years earlier by Grote-Hynes \cite{rfg-jth} and
H\"anggi-Mojtabai \cite{ph-fm}.

Third, we consider the case where the correlation time of
all the noise processes (internal and external) are vanishingly small,
\begin{eqnarray*}
\langle f(t) f(t^{\prime}) \rangle &=& 2 \gamma k_B T
\delta(t-t^{\prime}), \\
\langle \xi(t) \xi(t^{\prime}) \rangle &=& 2 D_R \delta(t-t^{\prime})
\end{eqnarray*}

\noindent where $D_R$ is the external noise strength in the limit
$\tau_R \rightarrow 0$ [see Eq.(\ref{eq5})] and $\gamma$ is the
dissipation due to internal thermal noise processes. In such a case,
we have
\begin{eqnarray*}
&& D_0 = D_R = k_B T + \frac {D_R}{\gamma},
\psi_0 = \psi_b =0, \\
&& \mu =- ( a_{-} + \gamma) \text{ and } a_{-} = -\frac{\gamma}{2} -
\left ( \frac {\gamma^2}{4} + \omega_b^2 \right )^{1/2}
\end{eqnarray*}

\noindent and hence the rate becomes
\begin{equation}\label{eq51}
k = \frac{\omega_0}{2 \pi \omega_b} \left [ \left ( \frac {\gamma^2}
{4} + \omega_b^2 \right )^{1/2}  - \frac{\gamma}{2} \right ] \exp
\left ( -\frac{\gamma E_b}{\gamma k_B T + D_R} \right ).
\end{equation}

\noindent In the limit $D_R \rightarrow 0$ (i.e., $D_{\epsilon}
\rightarrow 0 $ and  $D_{\pi} \rightarrow 0 $ simultaneously) we
recover the Kramers' results (\ref{eq48}). Also, the degree of
correlation $\lambda$ is implicitly present in the expression of
$D_R$ [see Eq.(\ref{eq5})]. In the expression (\ref{eq51}) in
addition to $T$, $D_R / (\gamma k_B)$ defines a new effective
temperature due to external driving. In a different context where
the heat bath is modulated by an external fluctuating field we have
also encountered the appearance of the effective temperature
\cite{jrc1,jrc2}.

Finally, to study the effect of degree of correlation of external
noise on the generalized escape rate (\ref{eq46}), we consider the
case where the fluctuation in the dynamics is only due to external
source, i.e $f(t) = 0 $. The statistical properties of the
correlated external noises are given by Eq.(3). Since, in this case,
the dissipation is independent of the fluctuations, we may assume
Markovian relaxation so that $\gamma(t) = \gamma \delta(t)$. We then
obtain after some lengthy algebra from the general expression
(\ref{eq46})
\begin{eqnarray*}
&& \phi_0 = \frac{D_R}{1+ \gamma \tau_R + \omega_0^2 \tau_R^2},
\phi_b \frac{D_R}{1+ \gamma \tau_R - \omega_b^2 \tau_R^2}, \\
&& \psi_0 = \frac{D_R \tau_R}{1+ \gamma \tau_R + \omega_0^2
\tau_R^2},
\psi_0 = \frac{D_R \tau_R}{1+ \gamma \tau_R - \omega_b^2 \tau_R^2}, \\
&& \mu = - \gamma - (1+\gamma \tau_R) a_{-}, \\
&& a_{-} = \frac{1}{(1+\gamma \tau_R)} \left [ - \frac{\gamma}{2} -
\left ( \frac{\gamma^2} {4} + \omega_b^2 \right )^{1/2} \right ]
\end{eqnarray*}

\noindent where $D_R$ and $\tau_R$ are given by (\ref{eq5}) and
(\ref{eq6}) respectively. Consequently, the rate becomes:
\begin{eqnarray}
k & = & \frac{\omega_0}{2 \pi \omega_b} \left ( \frac{1+ \gamma
\tau_R + \omega_0^2 \tau_R^2} {1+ \gamma \tau_R - \omega_b^2
\tau_R^2} \right )^{1/2} \left [ \left ( \frac {\gamma^2} {4} +
\omega_b^2 \right )^{1/2} -
\frac{\gamma}{2} \right ] \nonumber \\
&& \times \exp \left ( -\frac{\gamma (1+ \gamma \tau_R - \omega_b^2
\tau_R^2) }{D_R(1+\gamma \tau_R)} E \right ). \label{eq50}
\end{eqnarray}

\noindent It is interesting to note that the expression (\ref{eq50})
denotes the escape rate induced by pure correlated external
fluctuations, which apart from depending on the strengths and
correlation times of the noise processes, depends crucially on the
degree of correlation of the external fluctuations through the
parameters $D_R$ and $\tau_R$. The absence of thermal temperature
and the appearance of the dissipation constant $\gamma$ demonstrate
the non-thermal origin of the noise processes as well as the absence
of the FDR in the dynamics.

\begin{figure}[!t]
\includegraphics[width=1.0\linewidth,angle=0]{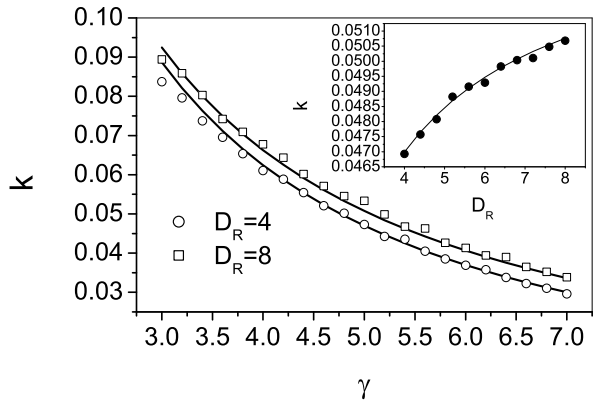}
\caption{\label{fig2} Barrier crossing rate $k$ as a function of
dissipation constant $\gamma$ for two different values of effective
noise strength $D_R$. The solid lines are drawn from theoretical
expression, Eq.(\ref{eq50}) and the symbols are due to numerical
simulation of Eq.(\ref{eq1}). The values of the parameters used are
$D_e=D_{\pi}=2$, $\tau_e=\tau_{\pi}=\tau=1$, $\lambda=0$ (open
circle) and $\lambda=1$ (open square). Inset: $k$ as a function of
$D_R$ for $\gamma=5$ and $0 \leqslant \lambda \leqslant 1$. Other
parameters are same as in the main figure.}
\end{figure}

\section{Numerical Implementation} \label{application}

To judge the potentiality and applicability of our recently
developed method for computation of the rate of barrier crossing
process, we discuss here both the details of the working equations
from the point of view of the numerical implementation and the
corresponding simulation of our method. To check the validity and
applicability of our model from the point of view of computational
implementation, we consider the dynamics in a bistable potential
$V(x)=(x^4/4)-(x^2/2)$ so that the activation energy becomes
$E=1/4$. We then numerically solve the Langevin equation (\ref{eq1})
by employing stochastic Heun's algorithm \cite{heun1,heun2}. The
numerical rate has been defined as the inverse of mean first passage
time \cite{jrc1,jrc2,mfpt} and has been calculated by averaging over
10,000 trajectories. In our simulation we have always used
$\tau_{\epsilon}=\tau_{\pi}=\tau=1$ such that the effective
correlation time $\tau_R$ is always equals to 1, independent of
values of the other parameters.  To ensure the stability of our
simulation, we have used a small integration time step $\Delta t =
0.001$ so that $\Delta t/\tau_R \ll 1$. To compare our theoretical
prediction with numerical simulation we consider the case where the
system is solely driven by an external correlated colored noise [see
Eq.(\ref{eq50})]. As the degree of correlation between the two noise
processes is increased the strength of effective noise strength
$D_R$ increases (see Eq.(\ref{eq5})). This effectively pumps more
energy into the system through which the escape rate should
increase. In Fig.(1) we see this effect clearly where the escape
rate $k$ is plotted as a function of the dissipation constant
$\gamma$ (in the limit of moderate to large friction regime) for two
different values of effective noise strength $D_R$ which are
evaluated by using two extreme values of the degree of correlation
$\lambda$ (0 and 1). In the inset we show more explicitly how the
system receives more energy through $D_R$ as the degree of
correlation increases. This behavior suggests that in a properly
designed experiment one can enhance the escape rate by externally
controlling the degree of correlation between the external
fluctuations.

\section{Conclusions}\label{conclusion}

In this paper, we have generalized the Kramers' theory of activated
rate processes for non-equilibrium open system where the system is
driven by two external cross-correlated noise processes with the
assumption that the underlying dynamics is non-Markovian. The theory
takes into account both the external  and internal fluctuations in a
unified way. The external fluctuations considered are stationary,
Gaussian. Our treatment is valid for intermediate to strong damping
limit. We have shown that not only the motion at the barrier top is
influenced by the cross correlation between the external
fluctuations, it has an important role to play in establishing the
stationary state near the bottom of the source well for open system.
The stationary distribution function in the well depends
significantly on the degree of correlation of the external noise
processes. We then derived the generalized Kramers' rate for the
open system and examined several limiting cases. To establish the
applicability and potentiality of our recently developed method, we
then numerically simulated the dynamics in a model bistable
potential and compared with one of the limiting cases. Our results
generated via numerical simulation reflect a good agreement with the
corresponding values obtained analytically. Our numerical analysis
clearly depicts that the escape rate can be enhanced by increasing
the degree of correlation between the external fluctuations.

\begin{acknowledgments}
JRC and SC would like to acknowledge the UGC, Delhi [MRP
Scheme-2007] for financial support. SKB acknowledges financial
support from Department of Physics, Virginia Tech.
\end{acknowledgments}


\begin{thebibliography}{99}
\bibitem{rdf} The reduced distribution function is defined as
$\tilde{P}_{st}(x)=\int_{-\infty}^{+\infty} P_{st} (x,v)$.
\end{thebibliography}
\end{document}